# Voltage-embedded equivariant machine learning potential for open system simulations

Yiming Guan[1], Rongzhi Gao[1,2,3*], ChiYung Yam[2,4], GuanHua Chen[1,2,*], Ziyang Hu[1,2,*]

[1]Department of Chemistry, The University of Hong Kong, Pokfulam, Hong Kong SAR, China
[2]Hong Kong Quantum AI Lab Limited, Pak Shek Kok, Hong Kong SAR, China
[3]MattVerse Limited, Pak Shek Kok, Hong Kong SAR, China
[4]Shenzhen Institute of Advanced Study, University of Electronic Science and Technology of China, Shenzhen, China

[*]Corresponding authors
Email: rzgao@yangtze.hku.hk, ghc@everest.hku.hk, hzy@yangtze.hku.hk

## Abstract

Modeling electrochemical interfaces under operational non-equilibrium conditions is vital for energy technologies but remains bottlenecked by the expensive cost of *ab initio* methods. Current machine learning potentials, largely designed for closed systems under homogeneous electric fields, are limited in open quantum transport applications. To overcome this, we present an $E(3)$-equivariant graph neural network that embeds voltage bias for open-system simulations. Our approach decouples the system energy and forces into zero-bias and bias-dependent contributions, assigning distinct vector encodings to electrode and scattering-region atoms to capture non-equilibrium conditions. Trained on limited discrete bias data, the model achieves high predictive accuracy and robust extrapolation transferability. When applied to a lithium/water interface, our model successfully captures the field-induced dynamic reorientation of water molecules and reproduces asymmetric electrochemical behavior at different electrodes.

## Introduction

Electrochemical interfaces (ECIs) serve as the fundamental site for energy conversion and storage, playing a pivotal role in critical technologies such as batteries, fuel cells, and electrocatalysis.[1-3] These heterogeneous boundaries are where charge transfer occurs and chemical bonds are forged or broken, dictating the efficiency and longevity of devices.[4, 5] Moreover, realistic electrochemical interfaces operate under non-equilibrium conditions driven by an applied bias voltage, which breaks the system symmetry and induces a continuous flow of charge carriers, creating a dynamic environment that fundamentally differs from an isolated system at equilibrium.[6-10]

To address this problem, DFT combined with nonequilibrium Green's functions (NEGF) serves as an effective method to model quantum transport under external bias voltages which can be used to simulate the electrode-electrolyte interactions and applied in the field like solid-electrolyte interphase to better comprehend its electronic insulating properties and formation mechanisms.[11-15] Consequently, explicitly including the bias is not merely a refinement but a necessity for bridging the gap between theoretical models and operational devices.

However, traditional *ab initio* methods suffer expensive computational cost, especially in the NEGF formalism, treating the open boundaries requires computationally intensive integration of Green's functions over energy.[16-22] These factors make it unfeasible to perform long-time molecular dynamics (MD) simulations for large-scale interface models, which are essential for understanding realistic device behavior.[23-26] To overcome this bottleneck, machine learning potentials (MLPs) have emerged as a promising solution.[27-32] Leveraging limited reference datasets obtained from quantum mechanical calculations, MLPs can accelerate MD simulations by orders of magnitude while retaining *ab initio* accuracy, thus enabling the study of complex interfacial dynamics.[33-38] Besides, for electrical effects, significant progress has also been made in different MLPs, which treat the electric field as a global vector feature to interact with atomic local feature representations, allowing for the accurate prediction of polarization, dielectric response and spectra in isolated molecules.[39-42]

Nonetheless, these MLPs responding to electric effects are theoretically designed for closed systems under a homogeneous electric field, which poses a critical limitation when applied to open quantum transport systems modeled by NEGF.[40, 42, 43] Explicitly including dipole and polarizability terms as input features or in the loss function constrains the model to a second-order approximation of the electrostatic response.[39, 40, 44] Moreover, this truncation is particularly problematic for interfacial environments under an applied bias. Interfacial systems are distinct from bulk phases due to their inherent non-uniformity and the presence of non-equilibrium charge injection.[45, 46] Crucially, the bias voltage in NEGF-based description is fundamentally different from the external electric field. For the applied voltage, it acts as a boundary condition that establishes a difference in electrochemical potential between two macroscopic reservoirs.[47-49] This potential difference drives a continuous, non-equilibrium injection of charge carriers into the scattering region, resulting in a complex, spatially varying potential drop that depends intimately on the atomic configuration and the screening properties of the interface. Consequently, conventional MLPs designed for external electric fields are insufficient for electrochemical interfaces to capture the intricate charge redistribution and the distinct boundary conditions in realistic transport simulations.

In this work, we propose a method that integrates bias voltage into an MLP for open-system simulation by splitting the energies and forces of model systems into chemical and voltage contributions. We demonstrate the capability of the NEGF-based framework by applying it to a lithium-lead water system, enabling a comprehensive analysis of both the electric response of bulk water in the central region and the specific ion hydration structure of lithium at the electrode interface. Moreover, we reveal that the voltage contribution can be learned robustly from limited training data. Even when trained on a sparse set of discrete bias, the model achieves high predictive accuracy for unseen voltages in both interpolation and extrapolation regimes.

## Results

**Voltage embedding in machine learning potential.** To accurately describe systems under a given voltage bias, we select the $E(3)$-equivariant graph neural network (GNN) architecture as the basic model.[50-52] However, standard GNN potentials are typically designed for closed systems and lack an intrinsic mechanism to account for the effects of bias voltage. To overcome this limitation, we propose a two-component modeling strategy that decouples the system energy into 0V-based contribution $E_0$ and a bias-dependent correction $E_\mathrm{V}$:

$$E_\mathrm{total}(\mathbf{R},\mathbf{Z},V) = E_0(\mathbf{R},\mathbf{Z}) + E_\mathrm{V}(\mathbf{R},\mathbf{Z},V),$$

where $\mathbf{R}$, $\mathbf{Z}$, and $V$ are atomic coordinates, atomic numbers, and bias voltage, respectively. $E_0$ is the potential energy at 0 V representing the intrinsic chemical contribution, while $E_\mathrm{V}$ is the bias-related energy varied with the value of applied voltages.

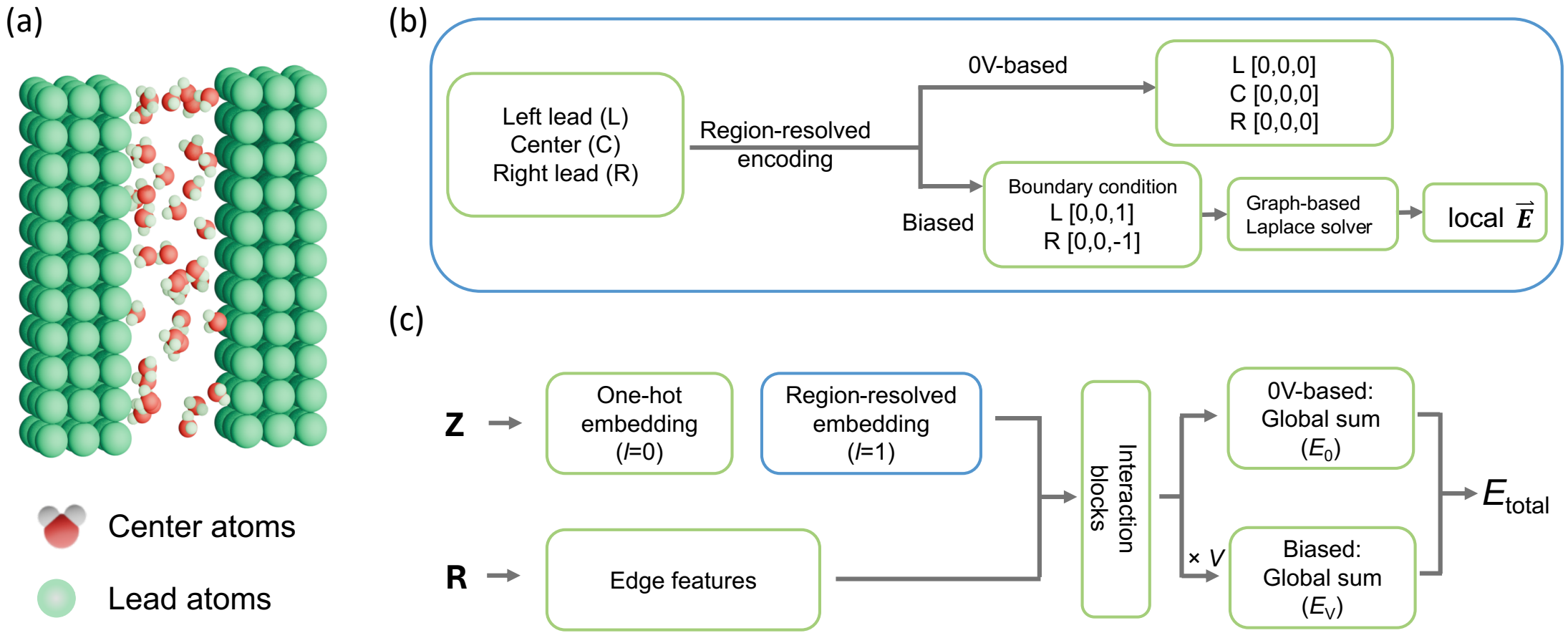

**Figure 1 The architecture of the model. a**, Schematic representation of the device structure, partitioned into the left lead, central scattering region (center atoms), and right lead. **b**, Region-resolved encoding scheme designed to incorporate non-equilibrium boundary conditions. Atoms are assigned distinct embedding vectors based on their physical location—left lead (L), center (C), or right lead (R). The 0V-based model employs zero vectors for all regions, whereas the biased configuration assigns directional region-resolved unit vectors aligned with the transport axis ([0, 0, 1] and [0, 0, −1]) to the respective leads, which are further sent to a discrete graph-based Laplace solver to get the atomic local electric field. **c**, The total energy $E_\mathrm{total}$ is decoupled into 0V-based contribution ($E_0$) and a bias-dependent correction ($E_\mathrm{V}$). Initial node features are constructed from atomic numbers ($\mathbf{Z}$) via one-hot embedding ($l$=0) and updated with the region-resolved encoding ($l$=1), while atomic coordinates ($\mathbf{R}$) are used to compute structural edge features. Following feature refinement through shared interaction blocks, the 0V-based pathway aggregates per-atom outputs via global sum to yield $E_0$. Concurrently, the biased pathway scales the extracted features by the scalar applied biased voltage ($V$) before global summation to compute the correction term $E_\mathrm{V}$.

As shown in **Figure 1**, the atoms are categorized into three types, each belonging to left leads, central region and right leads, respectively. Similar with conventional equivariant MLPs, the atomic number **Z** is used to construct the initial irreducible representations with rotation order ($l$) of 0. Besides, to incorporate the non-equilibrium boundary conditions, a region-resolved unit vector, $\mathbf{v}_\mathrm{r}$, is assigned to each atom used to construct the higher order features with $l$ = 1. For 0V-based model, $\mathbf{v}_\mathrm{r}$ is [0, 0, 0] for all three regions. For biased model, the Hartree potential of the system establishes a near-linear drop across the central region, with the chemical potentials of the left lead ($\mu_L$) and right lead ($\mu_R$) asymptotically approaching $\mu_{L/R} = E_F \pm eV/2$, where $E_F$ is the Fermi levels. The given bias voltage ($V$) will lead to a directional electric field along transport axis in the central region. Here, under Dirichlet potential boundary conditions, we firstly set [0,0,1] for atoms in the left lead (source), [0,0,-1] for atoms in the right lead (drain) and then employ a discrete graph-based Laplace solver to propagate these boundary conditions into the central region, reconstructing the continuous electrostatic potential field across the entire device. Subsequently, atomic electric fields are calculated and mapped to the $\mathbf{v}_\mathrm{r}$ for all atoms. After the initial embeddings being established, the features are then processed through a series of interaction blocks, which perform equivariant message passing to learn complex many-body atomic environments. In the final readout block, the per-atom energies from the 0V-based model are summed to get $E_0$. For the biased model, the scalar magnitude of the bias voltage $V$ is multiplied with the final interaction block features to output $E_\mathrm{V}$. The details of model structure and training parameters can be found in **Supporting Information (**model architecture section**)**.

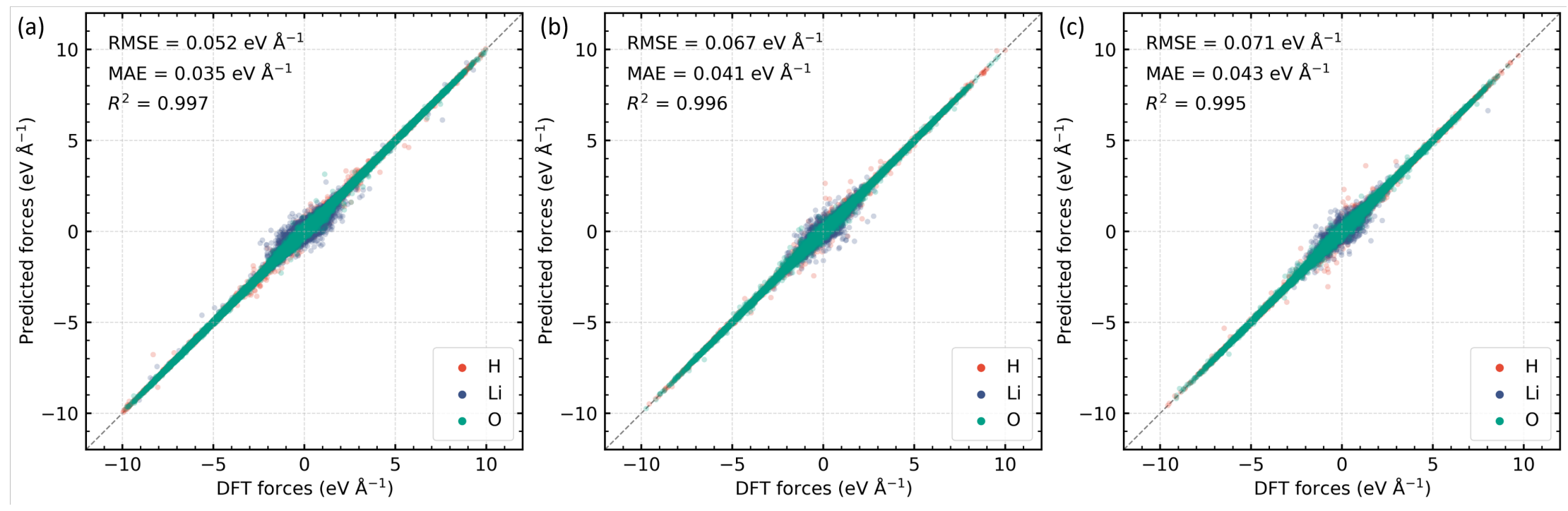

**Figure 2 Atomic forces prediction performance of the 0V-based model.** Atomic forces comparison of **a**, training set, **b**, validation set, and **c**, test set between 0V-based model predictions and reference DFT values.

**Li/water dataset and model performance.** To evaluate our approach, we employed CP2K and SMEAGOL in NEGF-DFT framework to investigate the solvation behavior of lithium atoms in water.[16, 53] The device structure can be divided into two identical body-centered cubic lithium (bcc-lithium) leads and a central scattering region (**Figure S1**). To mimic the bulk electrode properties, atoms in the lead regions are fixed at their equilibrium lattice positions. The central region comprises the water molecules and the adjacent lithium screening layers, while the dynamic interface is formed at the boundary between the electrode surface and the water molecules. 8969 devices structures are generated and calculated at 0 V (where $\mu_L = \mu_R$) in NEGF-DFT framework to construct the equilibrium 0V dataset, while the bias contribution containing the excess energies and forces are calculated to generate a delta dataset for biased model training, which are expressed by the following equations,

$$E_{\mathrm{V}} = E_{\mathrm{ref}} - E_0,$$
$$F_{\mathrm{V,i}} = F_{\mathrm{ref},i} - F_{0,i},$$

where $E_{\mathrm{ref}}$, $F_{\mathrm{ref},i}$ are the energy and atomic forces calculated by NEGF-DFT at specific bias voltage, and $E_0$, $F_{0,i}$ are those at 0 V. The sampling points for the bias voltage range from 0.25 to 1.50 V with a step size of 0.25 V.

We first assess the predicted forces at zero bias against the training, validation and test dataset with a splitting ratio of 8:1:1. As shown in **Figure 2**, for the 0V-based model, the atomic force RMSE and MAE are 0.052 eV Å$^{-1}$ and 0.035 eV Å$^{-1}$ for the training set, 0.07 eV Å$^{-1}$ and 0.04 eV Å$^{-1}$ for the validation and test sets. The directional force components of all three atom types are tightly around the diagonal identity line and align well with its determination coefficient ($R^2$) of more than 0.995. Furthermore, to assess the physical fidelity of the trained 0V-based model, we conducted NVT molecular dynamics simulations on water molecules and analyzed the structural properties via radial distribution functions (RDFs). As illustrated in **Figure S3**, the corresponding RDF profiles derived from the model exhibit remarkable consistency with those obtained from *ab initio* MD (AIMD) trajectories.

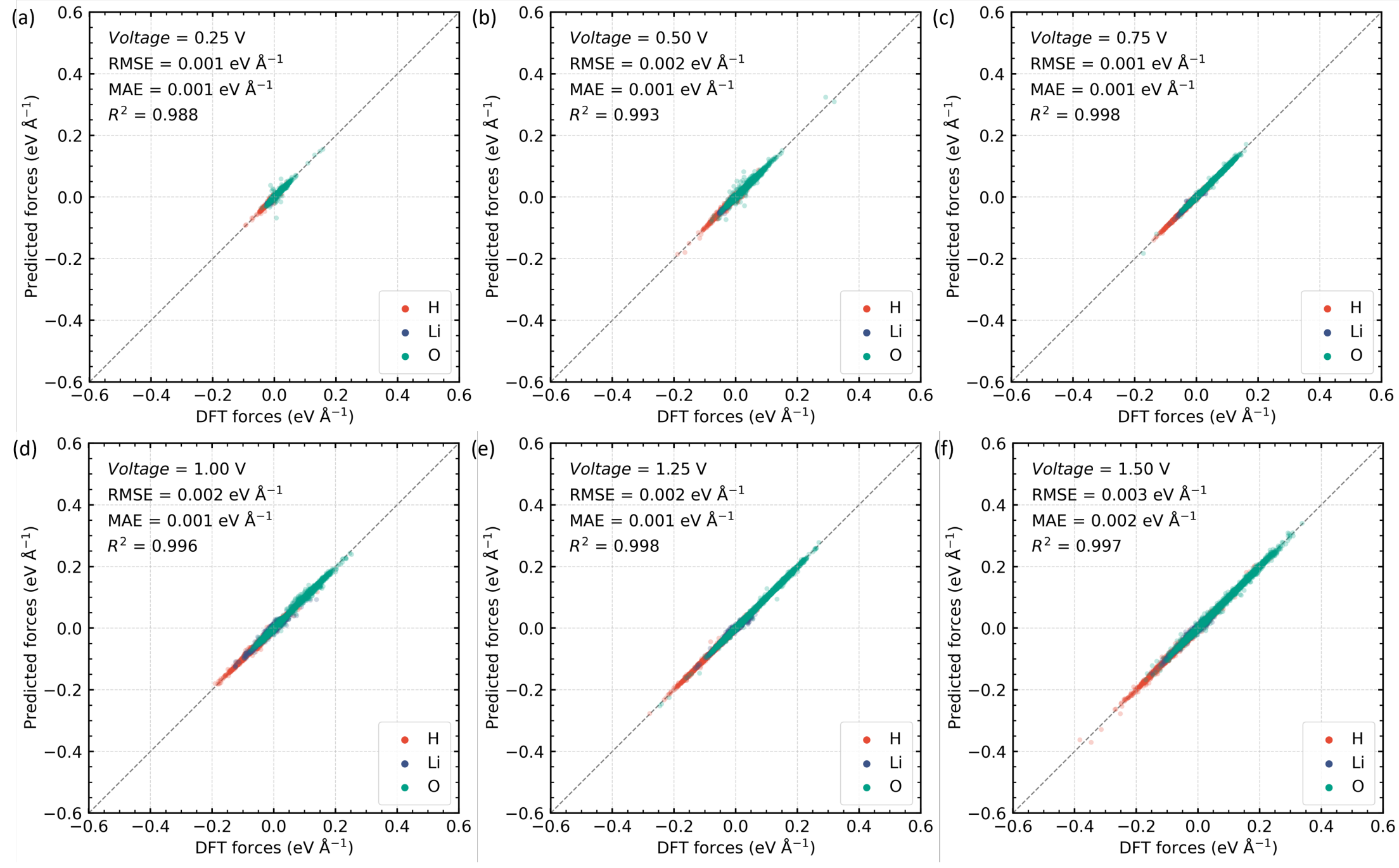

**Figure 3 Performance of the biased model in the training voltage range.** Parity plots showing the predicted versus DFT-calculated voltage-induced force contributions for bias voltages **a**, 0.25 V, **b**, 0.50 V, **c**, 0.75 V, **d**, 1.00 V, **e**, 1.25 V, and **f**, 1.50 V.

To evaluate the bias-dependent contribution, we trained the biased model on the delta dataset as described before. **Figure S4** shows the relationship between the average excess atomic force components and bias voltages. The maximum forces values increase with the applied voltages for all *x*, *y* and *z* components. For the transportation axis (*z*-axis), the average excess atomic forces increase linearly with increase of voltages from 0.25 V to 1.5 V (**Figure S5**). For the bias-contribution model, the RMSE values for all voltages are in a small range from 0.001 to 0.003 eV/Å (**Figure 3**). Moreover, to verify the interpolation and extrapolation ability of bias-contribution model, we select the voltage with 0.88 V as interpolated value, 1.80 V and 2.00 V as extrapolated values. **Figure 4** shows the RMSE, MAE values and $R^2$ coefficients of voltage at 0.88 V, 1.80 V, and 2.00 V respectively. The interpolation data (0.88 V) shows a high $R^2$ coefficient with 0.98, and an RMSE of 0.005 eV Å$^{-1}$, which are comparable with those of voltage values used in training. Moreover, the biased model shows a good extrapolation transferability for the voltage beyond the training dataset. For voltage at 1.80V and 2V, the model predicted forces keep nearly linear relationship with DFT calculated forces with the $R^2$ coefficient of 0.978 and force RMSE on the order of 0.01 eV Å$^{-1}$, implicating the ability of extrapolation prediction for biased model.

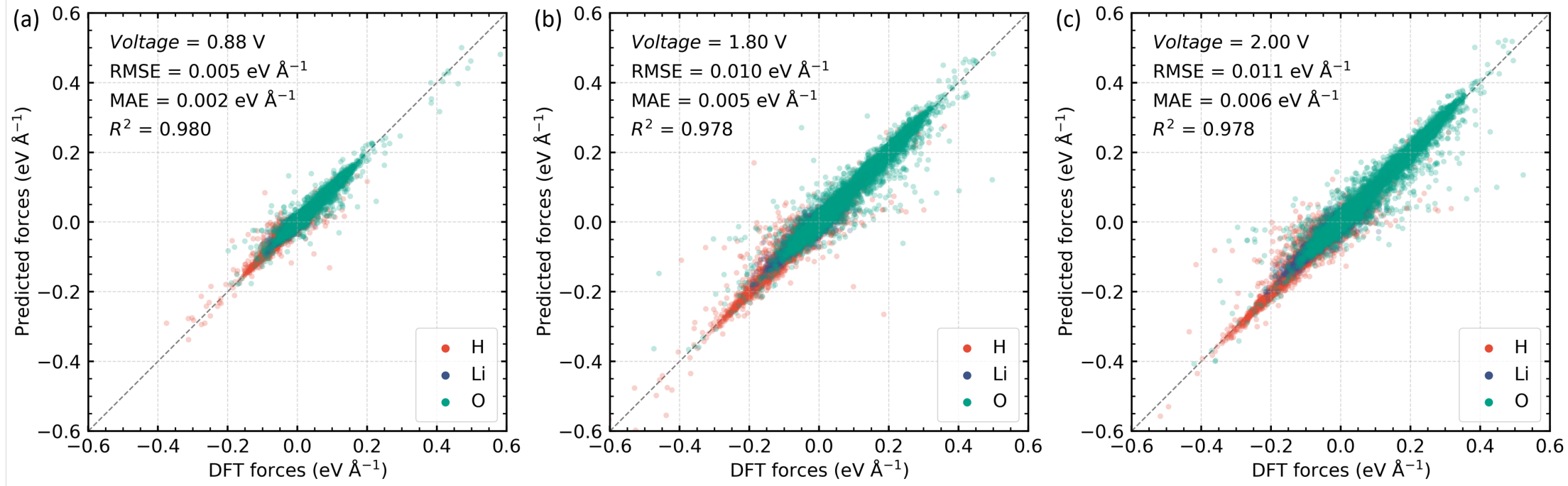

**Figure 4 Transferability of the biased model. a**, Interpolation performance at an unseen voltage of 0.88 V. Extrapolation performance at **b**, 1.80 V and **c**, 2.00 V.

**Field effect for bulk water.** In NEGF-DFT, the applied bias voltage induces an electrostatic potential difference between the right and left leads, resulting in an effective electric field along the transport direction. To elucidate the electrostatic potential distribution under non-equilibrium conditions, we analyzed the Hartree potential profiles calculated within the NEGF-DFT framework. **Figure S8** illustrates the voltage drop profile along the transport direction, defined as the difference between the Hartree potential $\Delta V_{\mathrm{H}}$ under finite bias $V_{\mathrm{bias}}$ and the equilibrium zero-bias potential $V_0$, which is calculated by:

$$\Delta V_{\mathrm{H}} = V_{\mathrm{bias}} - V_0.$$

It can be found that the potential drop is concentrated primarily within the central scattering region, while the potential in the electrode regions remains relatively flat due to effective metallic screening. Notably, within the bias range investigated (up to 2.0 V), the potential exhibits a distinct linear decay across the scattering region. Since the electric field is proportional to the spatial gradient of the potential, this linearity indicates that the internal electric field is effectively constant in this bulk region. Consequently, the effect of the applied bias on the central bulk region can be well-approximated as a uniform electric field and we further analyzed the water structuring response to voltage bias. As shown in **Figure 5c**, $\varphi$ is defined as the angle between water molecular dipole and field orientation, where the former is the bisector of the two intramolecular O-H bond vectors and the latter is the direction along the negative $z$ axis. **Figure 5a** depicts the orientation evolution in a 5-ps NEGF-AIMD simulation trajectory (T = 300 K) at 0 V, 0.5 V and 1.0 V, respectively. Under zero bias, $\varphi$ fluctuates around 85°–95°, indicating that the water molecules are in a disordered state dominated by thermal motion without voltage bias applied. Contrarily, upon applying a bias, a rapid declining change in $\varphi$ is observed within the first 2 ps, confirming the fast response of water molecules to the strong local electric field. We then use our model to perform NEGF-MD simulation (marked as ML-MD) under the same conditions. As illustrated in **Figure 5b**, the ML-MD trajectories at 0 V, 0.5 V, and 1.0 V exhibit consistency with the AIMD results in both the trends of the average values (solid lines) and the magnitude of fluctuations (shaded regions) of $\varphi$, signifying that the model has successfully captured the DFT-level interatomic forces. In addition, we run the ML-MD simulation at 2 V, where the bias value is not included in the training dataset. Compared with lower voltages, there is a stronger orientational alignment along the field direction, proving the transition from a thermally disordered state to a field-ordered state. To further evaluate the orientation distribution of water molecules at different voltages, we statistically analyzed the probability distribution of the angle as shown in **Figure 5c.** The bin width is set to 2° with the whole angle range from 0° to 180°, and the probability is defined as the number of angles in each bin width divided by the total angles number. It is worth noting that the angle $\varphi$ at 0V condition exhibits a broad plateau-like distribution with two shallow peaks around 60° and 120°, corresponding to the adsorption behavior at right and left active lithium leads (**Figure S10**). Upon applying the voltages, there is a narrowing and shifting of the angular peak towards 0°, which corresponds well with the orientation evolution, demonstrating the transition behavior well, which is also validated by previous studies.[40, 54, 55] **Figure 5d** depicts the mean, standard deviation (Std) and median values of $\varphi$. The mean value of $\varphi$ decreases from 91.4° to 31.6° with increasing voltages, while the Std value slightly decreases from 39.1° to 19.9° from 0V to 1V and then drops to only 15.4° at 2V, validating the strong orientational alignment at higher voltages.

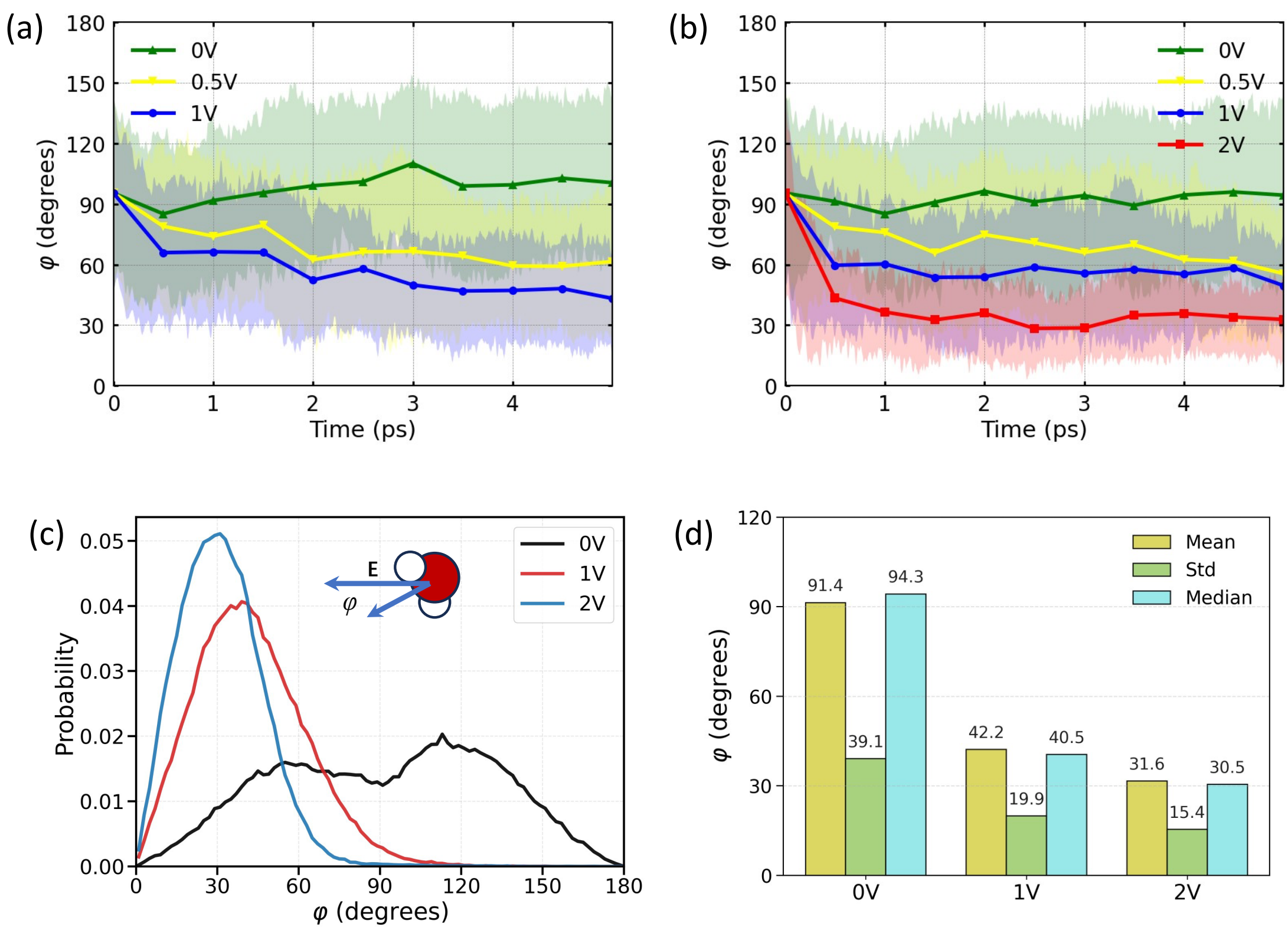

**Figure 5 Dynamic reorientation of water molecules under electric fields. (a-b)** Time evolution of the water dipole angle $\varphi$ (defined as the angle between the dipole moment and the negative *z*-axis) from **a,** AIMD and **b,** ML-MD simulations, respectively. **c,** Probability distribution of the angle $\varphi$ at 0

V, 1 V, and 2 V derived from ML-MD. **d,** Mean, Std and median values of angle $\varphi$ at 0 V, 1 V, and 2 V counted by water molecules from ML-MD.

We also present the radial distribution functions (RDFs) and vibrational density of states (VDOS) calculated from ML-MD trajectories to validate the effect of voltage bias on water structuring. In **Figure 6a**, the first peak of $g_{O\text{-}O}(r)$ shows a shifting trend towards the left, indicating a decrease in the O-O distance and a more compact packing under higher voltages bias. For the intramolecular O-H bond in **Figure 6b**, the increase in voltage induces a shift of the peak towards greater distance values, representing the elongated O-H bond length. **Figure 6c** depicts the intermolecular O-H distance distribution; the peak at ~1.8 Å becomes significantly sharper and shifts leftward, signaling a more structured and stronger hydrogen-bond network. All these changes can also be validated in the RDFs calculated from the AIMD trajectories shown in **Figure S11**.

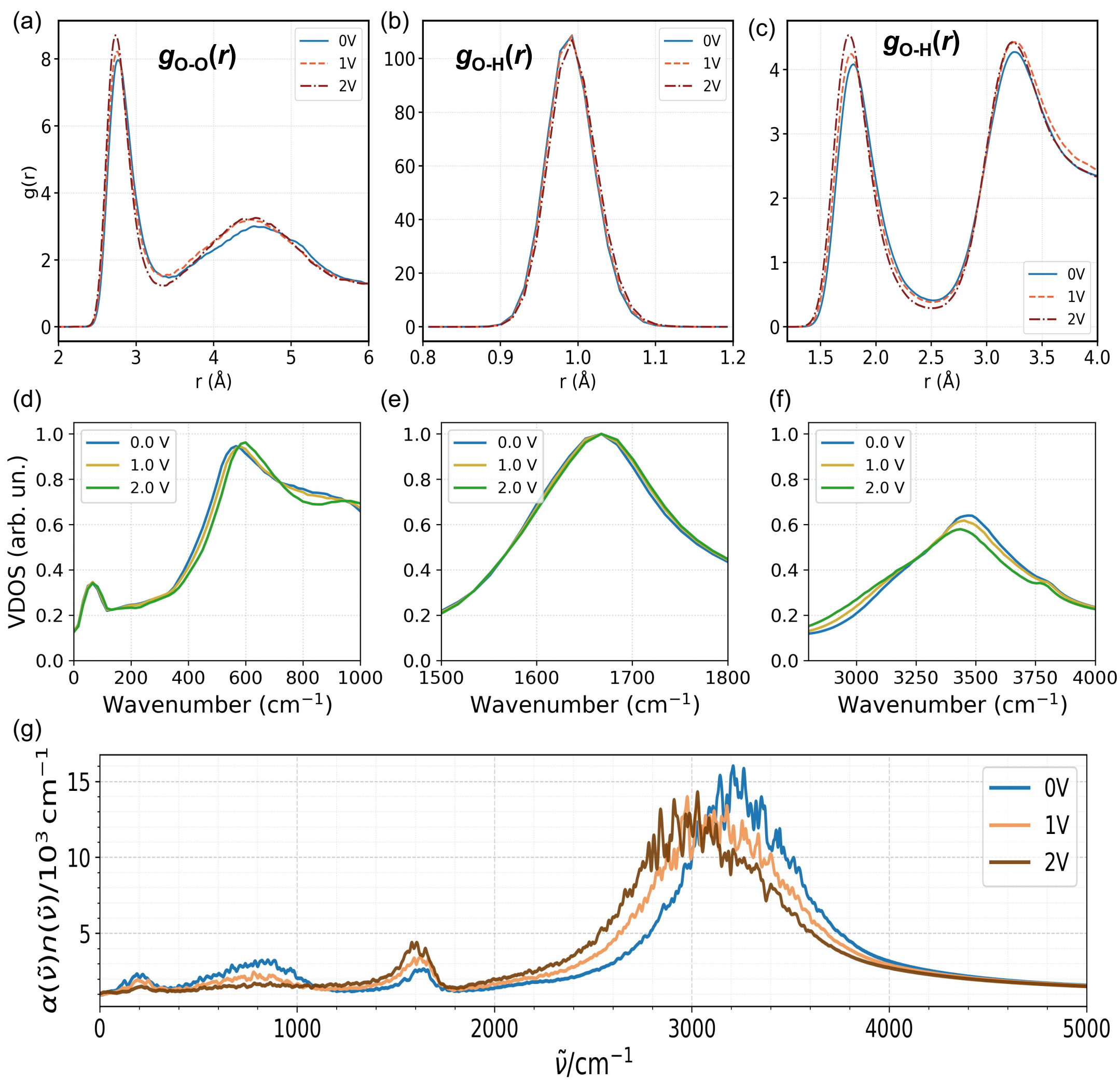

**Figure 6 Structural and dynamical characterizations of water under voltage bias calculated from ML-MD trajectories. (a-c)** Radial Distribution Functions (RDFs): **a,** The O-O RDF $g_{O\text{-}O}(r)$. **b,** The O-H RDF $g_{O\text{-}H}(r)$ ranging from 0.8 – 1.2 Å. **c,** The O-H RDF $g_{O\text{-}H}(r)$ ranging from 1.2 – 4 Å. **(d-f)**

Vibrational Density of States (VDOS): **d,** The low-frequency region (< 1000 $cm^{-1}$) dominated by translation and libration modes. **e,** The medium-frequency region (from 1500 – 1800 $cm^{-1}$) showing the H-O-H bending mode (~1650 $cm^{-1}$). **f,** The high-frequency region (from 2800 – 4000 $cm^{-1}$) showing the O-H stretching mode (> 3000 $cm^{-1}$). **g,** Comparison of IR spectra of water.

To further verify these structural observations from a dynamical perspective, we analyzed the VDOS profiles and IR spectra shown in **Figure 6d-g**.[56] Here a larger water box containing 674 water molecules is used to perform 200-ps ML-MD at different voltages. In the high-frequency region (**Figure 6f, g**), a significant red shift of the O-H stretching band (from ~3600 $cm^{-1}$ at 0V to ~3450 $cm^{-1}$ at 2V) is observed. This frequency reduction directly corresponds to the bond softening and elongation observed in the RDFs, a consequence of the vibrational Stark effect and enhanced hydrogen bonding. Conversely, the H-O-H bending mode in the mid-frequency range (**Figure 6e**,**g** ~1650 $cm^{-1}$) exhibits a slight blue shift in VDOS and substantial increasing intensity in IR spectra. The stiffening of the bending angle implies that water molecules are more strictly confined within the field-aligned network. In the low-frequency region (0-1000 $cm^{-1}$), it is noticed that the librational rotation mode is confined (**Figure 6d**), and the peak intensity (**Figure 6g**) gradually decreases and flattens out due to the field-induced dipole locking and strong interfacial adsorption. These dynamical fingerprints containing red-shifted stretching and blue-shifted bending align perfectly with the structural variations in RDFs, confirming the formation of a field-ordered, strongly hydrogen-bonded interface captured with high accuracy by our model.

**Voltage bias effect of the adsorption behavior of water molecules.** Finally, we focus on the lead-scattering region interface containing the lithium atoms and water molecules. In terms of different chemical potentials at left and right leads, they can be regarded as cathode and anode respectively. To better observe the hydration structure of lithium, we allow the screened lithium atoms to move and performed the ML-MD. As depicted in **Figure 7a-b**, the RDF of Li-O reveals an asymmetric response of the $Li^+$ hydration structure to the voltage biases at these two distinct interfaces. At the cathode (**Figure 7a**), the peak positioned at 1.98 Å refers to the first solvation shell of lithium atoms.[57] It is shown that this peak is gradually suppressed with increasing voltage, indicating the near-complete desolvation of the lithium atoms. On the contrary, the anode interface (**Figure 7b**) exhibits a significant enhancement and ordering of the lithium hydration structure. The first peak of the Li-O RDF becomes progressively higher and sharper with increasing voltage, signifying a more compact and well-defined first solvation shell. This asymmetric behavior is also supported by the calculated Li-O coordination numbers (CN) shown in **Figure S12**. When the voltage increases to 2 V, the CN at the cathode decreases to 0.18, while the CN at the anode increases to 1.82. Such phenomena can be further explained through the water orientation at two electrodes, as shown in **Figure S13a**, $\theta$ is defined as the angle between the water dipole moment vector and the Li-O bond direction. At the cathode (negative bias applied relative to the scattering region), the electric field vector points towards the electrode. Water dipoles align with this field, disrupting the optimal solvation shell geometry where oxygen lone pairs point towards $Li^+$. This field-dipole competition leads to the observed desolvation. Conversely, at the anode, the field alignment coincides with the hydration geometry, reinforcing the solvation shell. **Figure 7c-d** presents the potential-dependent interfacial water density profiles (box with 674 water molecules) as a function of the distance from the cathode and anode surfaces, respectively. A chemisorption density peak occurs symmetrically at 1.98 Å for both cathode and anode at 0V, which corresponds to the Li-O RDF first peak. Upon applying external voltage bias, a massive density enhancement is driven by the strong electrostatic attraction between the anode surface and the electronegative oxygen atoms of water. Conversely, the chemisorption peak at the cathode is severely suppressed, dropping to approximately 0.5 g/cm$^3$ at 1V and almost completely vanishing at 2V, and the water molecules are physically displaced outward, forming a restructured secondary peak at 3 Å.

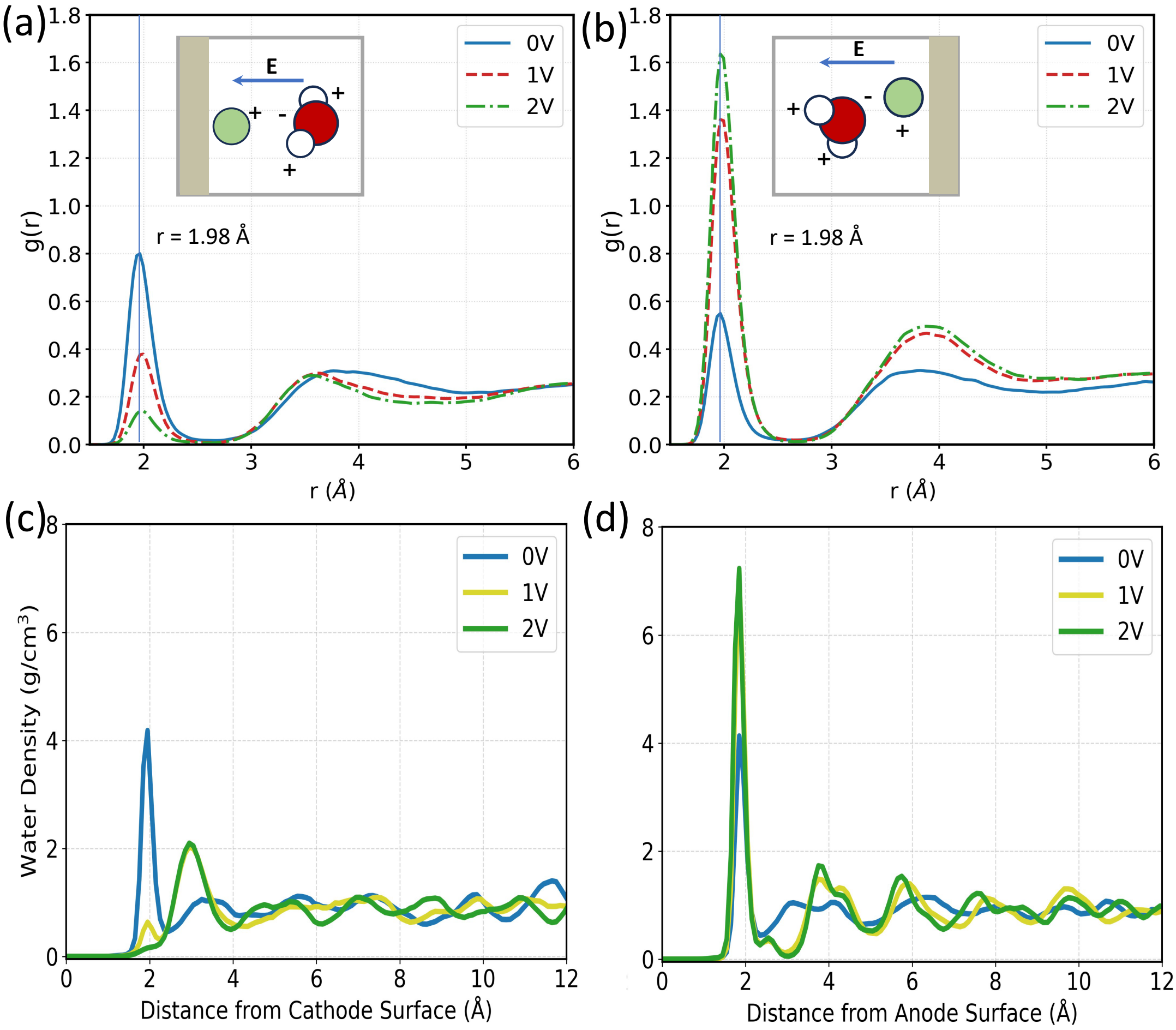

**Figure 7 Adsorption behavior of water molecules at electrode interfaces.** Li-O RDF at the **a**, cathode, and **b**, anode interface. Water density distribution profile in the z-axis direction starting from **c**, cathode and **d**, anode, respectively.

## Discussion

In this work, we present a machine learning framework which integrates an $E(3)$-equivariant neural network with voltage information encoded in NEGF-DFT. To enhance the generalizability, the energy and force contribution is divided into the zero-bias and bias-dependent terms. The usage of spherical harmonic irreducible representations of features and higher-order self-tensor products ensures the model to capture the complex quantum interactions. At the same time, by applying an additional vector encoding to the electrode atoms to represent the constant electron reservoirs of the left and right electrodes within the NEGF framework, and through multi-layer message passing, the model learns the nonlinear impact of the bias potential on the central regions. Remarkably, leveraging only a sparse sampling of bias voltages during training, it can accurately predict atomic forces under previously unseen voltages with a small RMSE of 0.011 eV/Å at 2 V, demonstrating the strong extrapolation capability and high efficiency in atomic force prediction. The model embedded with voltage bias under the NEGF framework can directly capture the electric field response of water molecules and reproduce the asymmetric electrochemical interfacial behavior of desolvation at the cathode and enhanced lithium hydration structure at the anode. In the future, we will focus on the complex non-equilibrium battery systems, especially the electronic insulating properties of solid electrolyte interphases.

## Data availability

The machine learning potential code is available via GitHub at https://github.com/guanyy18/NEGF-MLP. All data generated and analyzed during the study are available in the manuscript or can be obtained from the authors upon reasonable request.

## Acknowledgments

This work was supported by Hong Kong Quantum AI Lab Limited, Air @ InnoHK of Hong Kong Government, the GuangDong Basic and Applied Basic Research Foundation (Grant No. 2024A1515013283), the Guangdong Shenzhen Joint Key Fund (Grant No. 2019B1515120045), and the National Natural Science Foundation of China (Grant No. 22073007).

# Supporting Information

## Device Construction

Im3m lithium is selected as the lead,[1, 2] where its lattice parameter is set to 10.32 × 10.32 × 6.88 $Å^3$ as red rectangled in **Figure S1**. For better convergence, two extra lithium layers are included as buffers at both sides. For the central scattering region, we use a cubic water box whose size matches the lead's lattice parameter, and the number of water molecules is 33 to approximate the water density of 1 $g/cm^3$. The structure of water box is constructed by Packmol.[3] The transport direction is set to *z*. Besides, we scales the *c* length of the water box to generate more structures slightly deviating from the standard density of water.

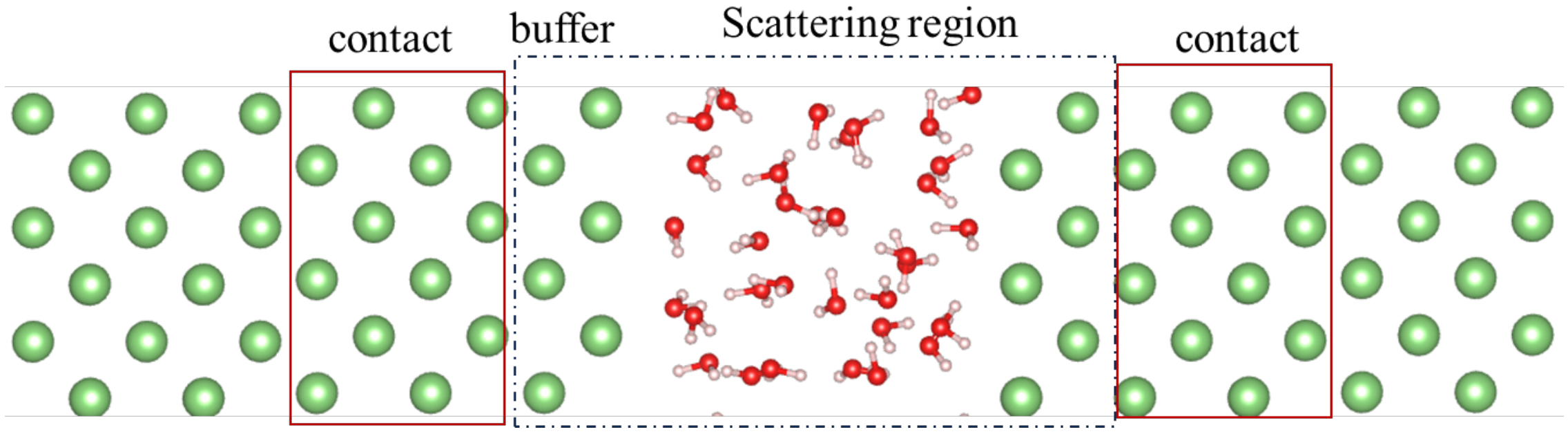

**Figure S2** A schematic diagram of the device structure.

## NEGF-DFT calculation

We use CP2K to perform the *ab initio* molecular dynamics (AIMD) simulations under the NEGF–DFT framework.[4-6] First we do the BulkTransport to calculate the self energy of leads. The PBE functional and an energy cutoff of 600 Ry are implemented.[7, 8] The Gaussian DZVP basis set (DZVP-MOLOPT-SR-GTH) is used for H and O, and the SZV basis set (SZV-MOLOPT-SR-GTH) for Li. The Gaussian and plane waves (GPW) methods with the Goedecker-Teter Hutter (GTH) pseudopotentials are used for all elements. A Monkhorst-Pack *k*-point grid of 1×1×20 is used for the bulk lead calculations. For the transport calculation, NVT ensemble with the temperature of 300 K and timestep of 0.5 fs are used to produce the atomic forces. The Grimme's D3 method is used to account for dispersion effects.[9] A Monkhorst-Pack *k*-point grid of 1×1×1 is used, and the SCF convergence criterion is $10^{-6}$ Hartree.

## Dataset Generation

To ensure a comprehensive sampling of the configurational space, we adopted a stochastic structure generation approach. For the first data collection, 5-ps AIMD segments were performed to obtain interfacial structures at 0V, which were then selected as a source to generate a large ensemble of uncorrelated structures by introducing random perturbations to the positions of atoms within the central scattering region. Specifically, Gaussian random displacements with a maximum amplitude of 0.5 Å were applied to the buffer atoms of lithium to mimic thermal disorder. For the water molecules, both random translational shifts and rotational reorientations were applied. Meanwhile, the bulk lead atoms remained fixed.

For the 0V-based and biased model, we constructed two datasets: 0V dataset and delta dataset, respectively. 8969 structures were generated for the 0V dataset, ensuring the structural diversity. For the delta dataset, we also run AIMD simulations (1500 steps with a timestep of 0.5fs) at various voltages (0.25 V, 0.5 V, 0.75 V, 1 V, 1.25 V, 1.5 V) to generate configurations and did the same perturbation operation for atoms in the central region. Besides, we selected structures from the 0-V dataset and did the NEGF–DFT calculation at the above voltages for all these generated structures. For each voltage value, more than 1000 structures are calculated, and the delta dataset contains total 7139 samples.

## Model Architecture

The atomic nodes encode the features of atomic numbers ($\mathbf{Z}$) and a region-specific unit vector ($\mathbf{v}_\mathrm{r}$) to encompass both the basic atomic information and external applied voltages (Figure S2 b). The intial atomic node can be expressed as:

$$h_{a,c}^{lp} = h_{a,c}^{l=0,p=1}(\mathbf{Z}) \oplus h_{a,c}^{l=1,p=-1}(\mathbf{v}_\mathrm{r})$$

where $l$, $p$, $a$ and $c$ are the rotation order, parity, atom index, and channel index. For the 0V-based model, $\mathbf{v}_\mathrm{r}$ = [0,0,0] is applied, whereas [0,0,1], [0,0,0] and [0,0, −1] are used for left lead, central region and right lead atoms in biased model.

The message-passing workflow of the interaction block is depicted in Figure S2 c, at each iteration, a new feature of the $i$-th atom is computed by aggregating information from its neighbors message. For the energy readout block, to utilize higher order information, we add the contribution of $l$ = 1 additionally,

$$E_0 = \sum_{i=1}^{N} \left[ E_\mathrm{atomic}(Z_i) + \alpha \left( \sum_{l=1}^{L-1} w_l^T h_i^{(l)} + \mathrm{MLP}_l \left( \left[ s_i^{(l)}; \left| v_i^{(l)} \right| \right] \right) \right) + \beta \right]$$

$$E_\mathrm{V} = \sum_{i=1}^{N} \left[ E_\mathrm{V,atomic}(Z_i) + \alpha \left( \sum_{l=1}^{L-1} w_l^T (h_i^{(l)} V) + \mathrm{MLP}_l \left( \left[ s_i^{(l)}; \left| v_i^{(l)} \right| \right] V \right) \right) + \beta \right]$$

where $E_\mathrm{atomic}(Z_i)$, $E_\mathrm{V,atomic}(Z_i)$ are the reference atomic energies for the 0V and delta training dataset, respectively. $V$ is the value of applied biased voltage and incorporated into the model by directly multiplying it with the features of each channel. For the last message-passing layer $L$, $s_i^{(l)}$, $\left| v_i^{(l)} \right|$ are the atomic features of $0e$ and the Euclidean norm features of $1o$, which are concatenated along the scalar feature dimension. $\alpha$, $\beta$ are the scaling and shifting parameters. The forces of atoms in the central region can be directly caluated using auto-differentiation as $\mathbf{F}_{i,c} = -\nabla_i E_\mathrm{total}$.

The total loss function is composed of energy and force parts,

$$\mathcal{L}_{total} = \lambda_E \cdot \mathcal{L}_{energy} + \lambda_F \cdot \mathcal{L}_{force}$$

where $\lambda_E$, $\lambda_F$ are the weights of energies and forces loss, respectively. The detailed energy and force loss terms are shown below,

$$\mathcal{L}_{\text{energy}} = \frac{1}{N_{\text{batch}}} \sum_{g=1}^{N_{\text{batch}}} \left\| E_g^{\text{pred}} - E_g^{\text{true}} \right\|^2$$

$$\mathcal{L}_{force} = \frac{1}{3N_{\text{central}}} \sum_{i\in\text{central}} \sum_{\alpha\in\{x,y,z\}} \left\| F_{i,\alpha}^{\text{pred}} - F_{i,\alpha}^{\text{true}} \right\|^2$$

The hyperparameters of the 0V-based and biased models are listed in **Table S1** , the hidden layer of both networks uses the irreducible representations of '128×0$e$ + 128×1$o$ + 128×2$e$' with the highest rotation order of $l_{\max}$=2, the cutoff radius is 5 Å. The many-body correlation order is 3 for the 0V-based model and 2 for the biased model, and the number of the interaction layers is 2 for the 0V-based model and 4 for the bias-contribution model. In the training process, the weighting factors of energy ($\lambda_E$) and forces ($\lambda_F$) is set as 1 $\text{eV}^{-1}$ and 1000 Å/eV, respectively.

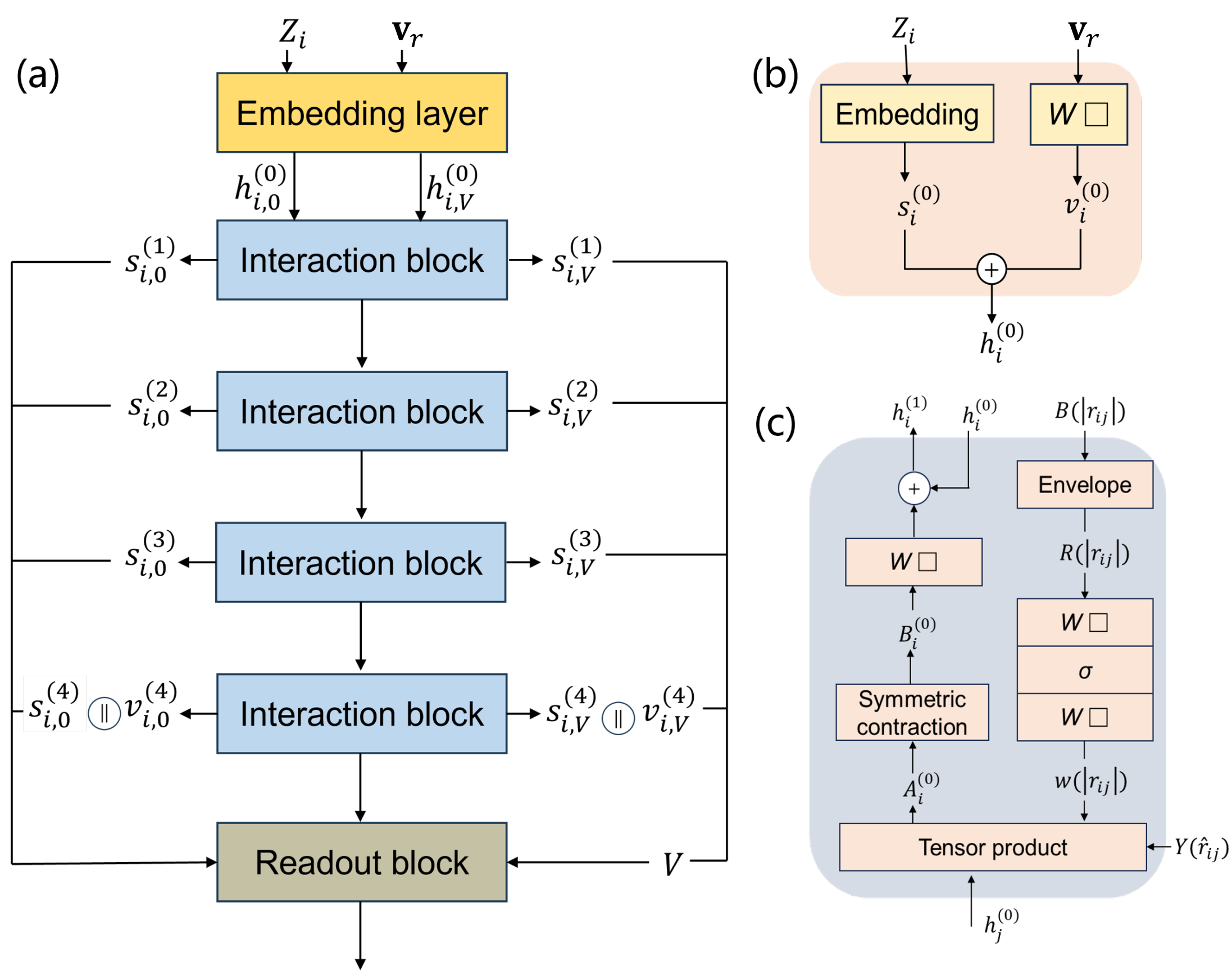

**Figure S3** Computational workflow of the voltage-embedded E(3)-equivariant neural network. **a,** Global information flow: The process initiates at the embedding layer where inputs $Z_i$ and $\mathrm{v}_r$ are transformed into equivariant features. These features propagate through the successive interaction blocks, where the information is split into 0V-based (left) and biased (right) streams. In the final stage, scalar and vector-norm features are concatenated and integrated with the external voltage $V$ in the readout block to generate the final output. **b,** Embedding layer workflow. **c,** Interaction block workflow.

**Table S1** hyperparameters of the 0V-based and bias-contrbution models.

| Model | Feature multiplicity | $l_{max}$ | Cutoff (Å) | Correlation number | Number of interaction layers |
|---|---|---|---|---|---|
| 0V-based | 128 | 2 | 5 | 3 | 2 |
| biased | 128 | 2 | 5 | 2 | 4 |

## ML-MD simulation

In this work, the machine learning molecular dynamics (ML-MD) were performed using the Atomic Simulation Environment (ASE) combined with trained 0V-based and biased model.[10] The NVT ensemble with the Nosé–Hoover thermostat was used in all simulations with a time step of 0.5 fs.[11] The lead atoms were fixed and the for each step, the energy and forces were computed by summing the outputs from two models. The simulation conditions were kept the same with the NEGF-AIMD simulations. To enhance sampling and improve statistical reliability, multiple independent ML-MD trajectories were initiated from the same initial structure but with different initial velocities. The initial velocities were randomly assigned to each atom from a Maxwell-Boltzmann distribution at the target temperature using different random seeds, ensuring statistical independence among trajectories. The Vibrational Density of States (VDOS) was computed via the Fourier transform of the velocity autocorrelation functions (VACFs) of the hydrogen atoms of water. Meanwhile, the infrared (IR) absorption spectra were obtained by the Fourier transform of the time correlation functions (TCFs) of the dipole moment time derivatives.[12] To establish a uniform spectroscopic probe across all molecular dynamics (MD) trajectories under varying running biases, the total dipole derivative $\dot{\boldsymbol{\mu}}(t)$ was evaluated on-the-fly as:

$$\dot{\boldsymbol{\mu}}(t) = \sum_{i \in \text{water}} \mathbf{F}_{\Delta,i}\,(V_{\text{ref}} = 1.0\,\text{V}) \cdot \mathbf{v}_i(t)$$

where $\mathbf{v}_i(t)$ represents the instantaneous atomic velocity, and $\mathbf{F}_{\Delta,i}(V_{\text{ref}} = 1.0\text{ V})$ denotes the voltage-dependent delta force evaluated at a constant reference bias of 1.0 V, acting as a consistent dynamical charge coefficient to probe the dipole fluctuations of the interfacial water under a standardized electric-field perturbation.

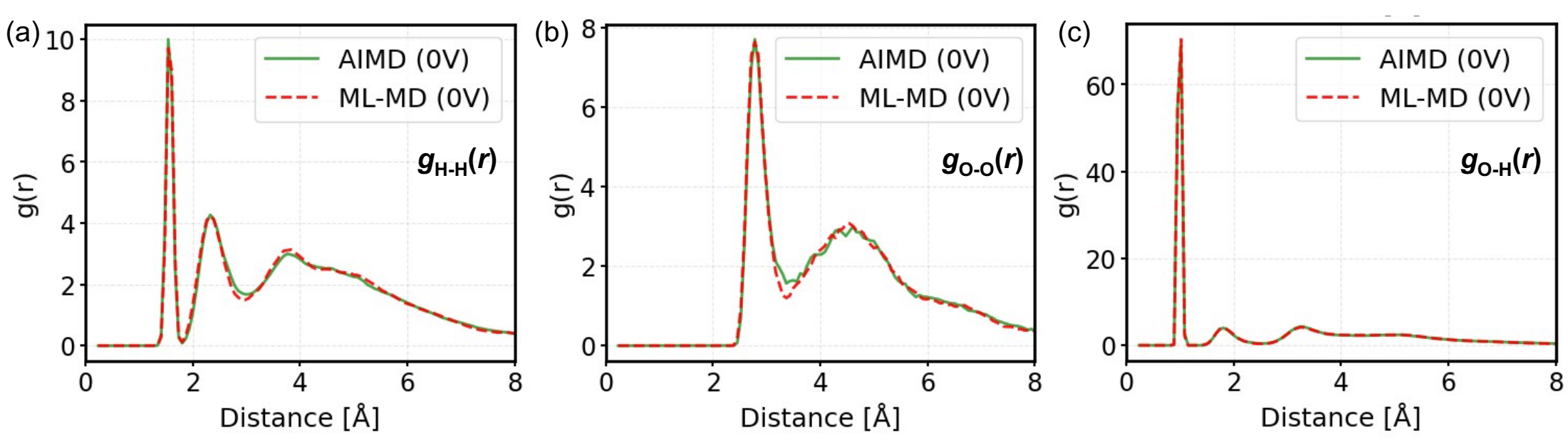

**Figure S4** Radial distribution functions of **a,** H–H, **b,** O–O and **c,** O–H generated by AIMD and ML-MD method at 0 V.

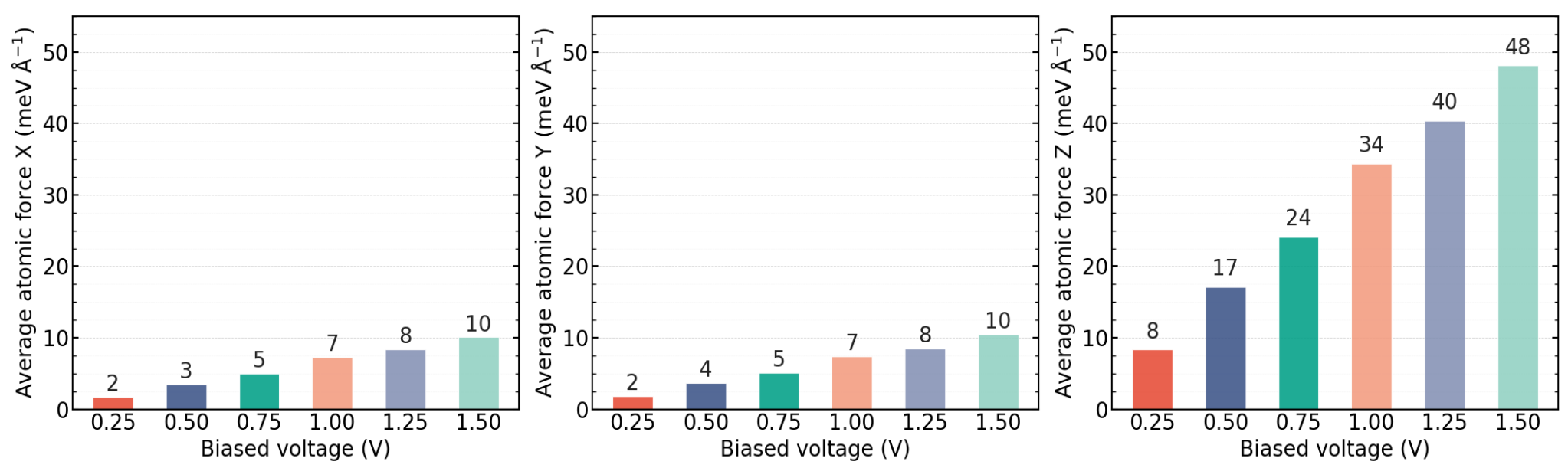

**Figure S5** Average absolute force components distribution in the delta dataset.

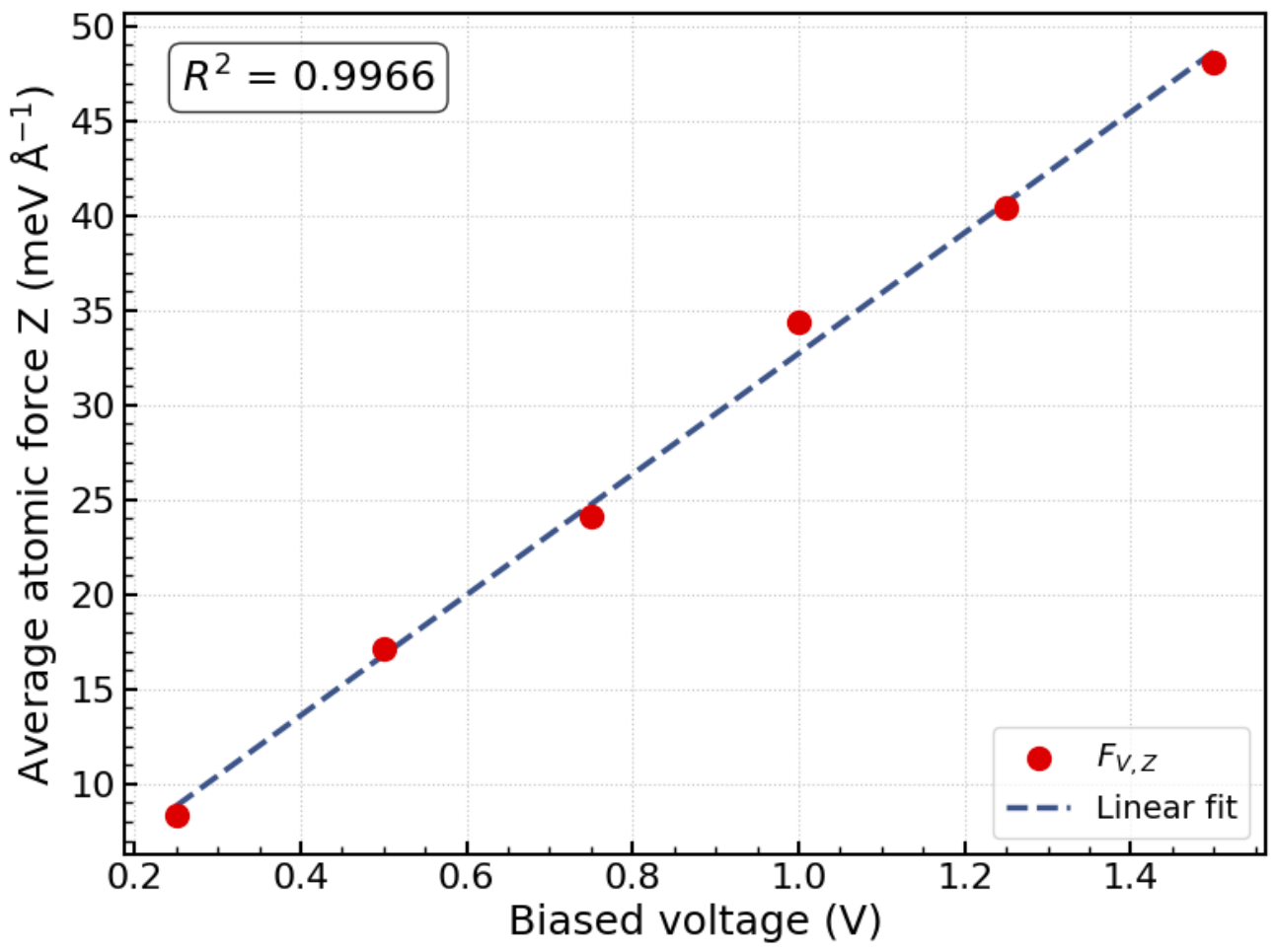

**Figure S6** Linear dependence of the average excess atomic force along the *z*-axis on the applied bias voltage. Red circles represent the mean absolute *z*-component of the atomic forces ($F_{V,z}$) in the delta dataset at various bias voltages. The blue dashed line indicates a linear fit to the data, demonstrating a strong linear correlation ($R^2 > 0.99$).

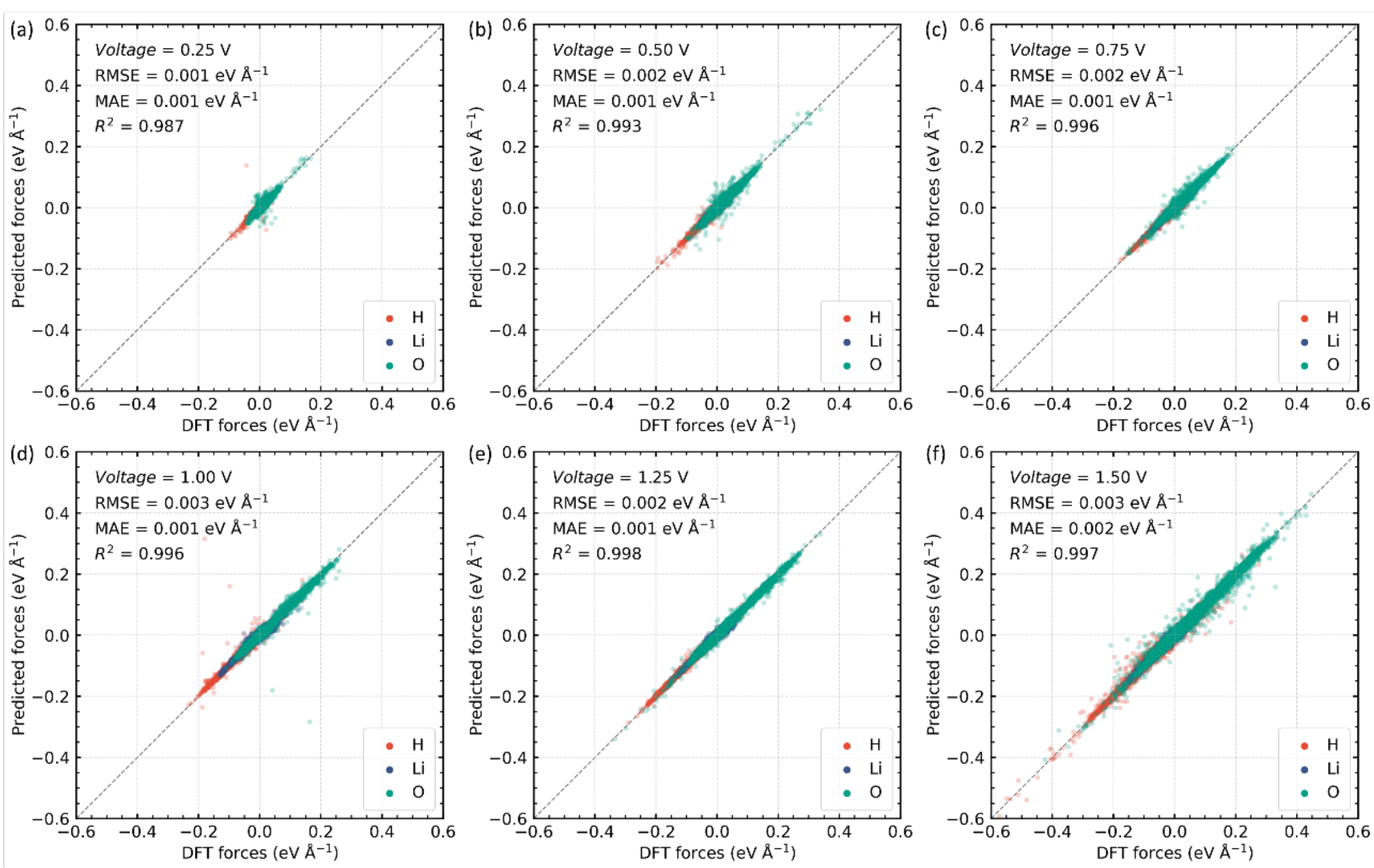

**Figure S7** Performance of the biased model in the training voltage range on training dataset. Parity plots showing the predicted versus DFT-calculated voltage-induced force contributions ($F_V$) for bias voltages ranging from 0.25 V to 1.50 V **(a-f)**.

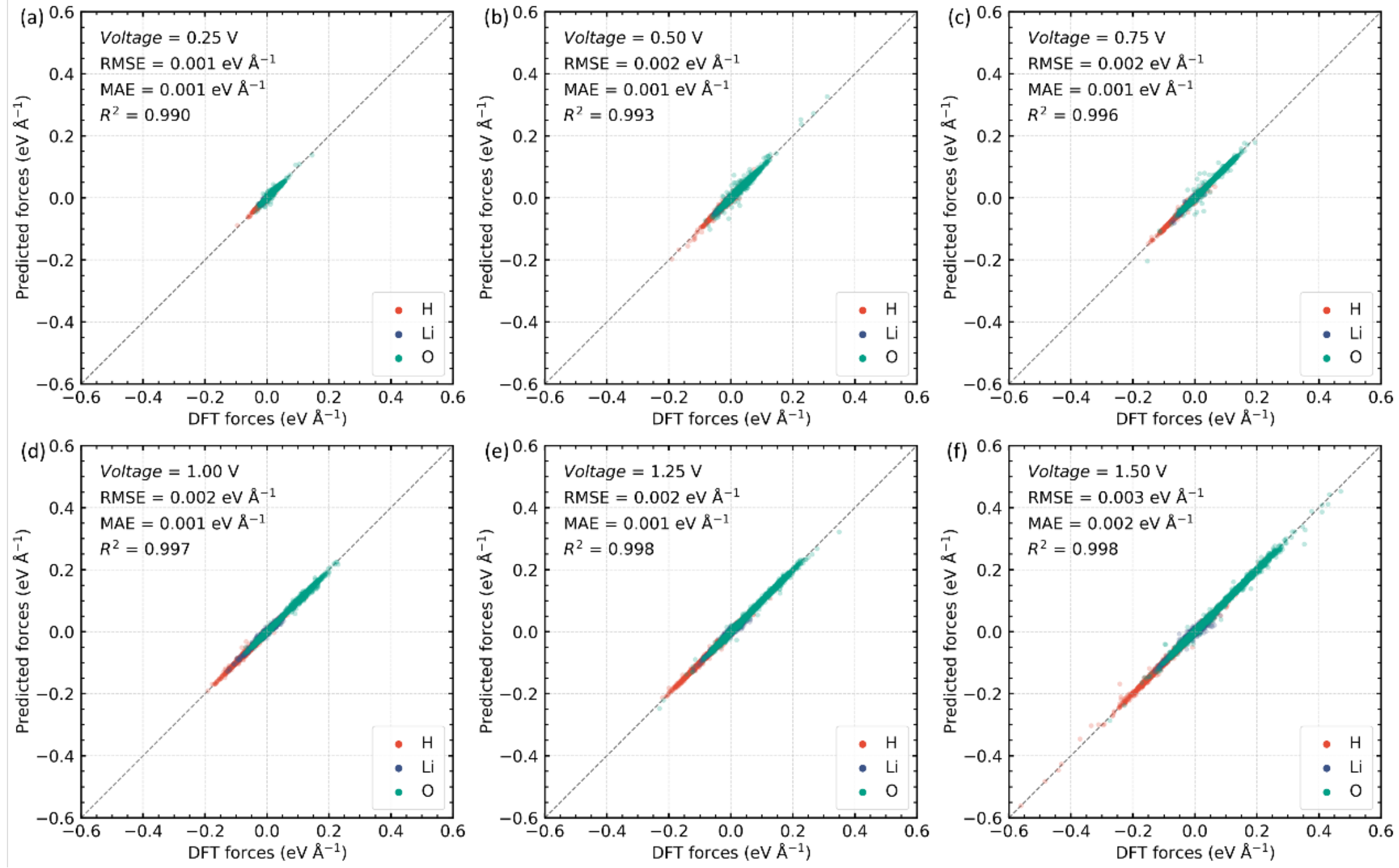

**Figure S8** Performance of the biased model in the training voltage range on validation dataset**.** Parity plots showing the predicted versus DFT-calculated voltage-induced force contributions ($F_V$) for bias voltages ranging from 0.25 V to 1.50 V **(a-f)**.

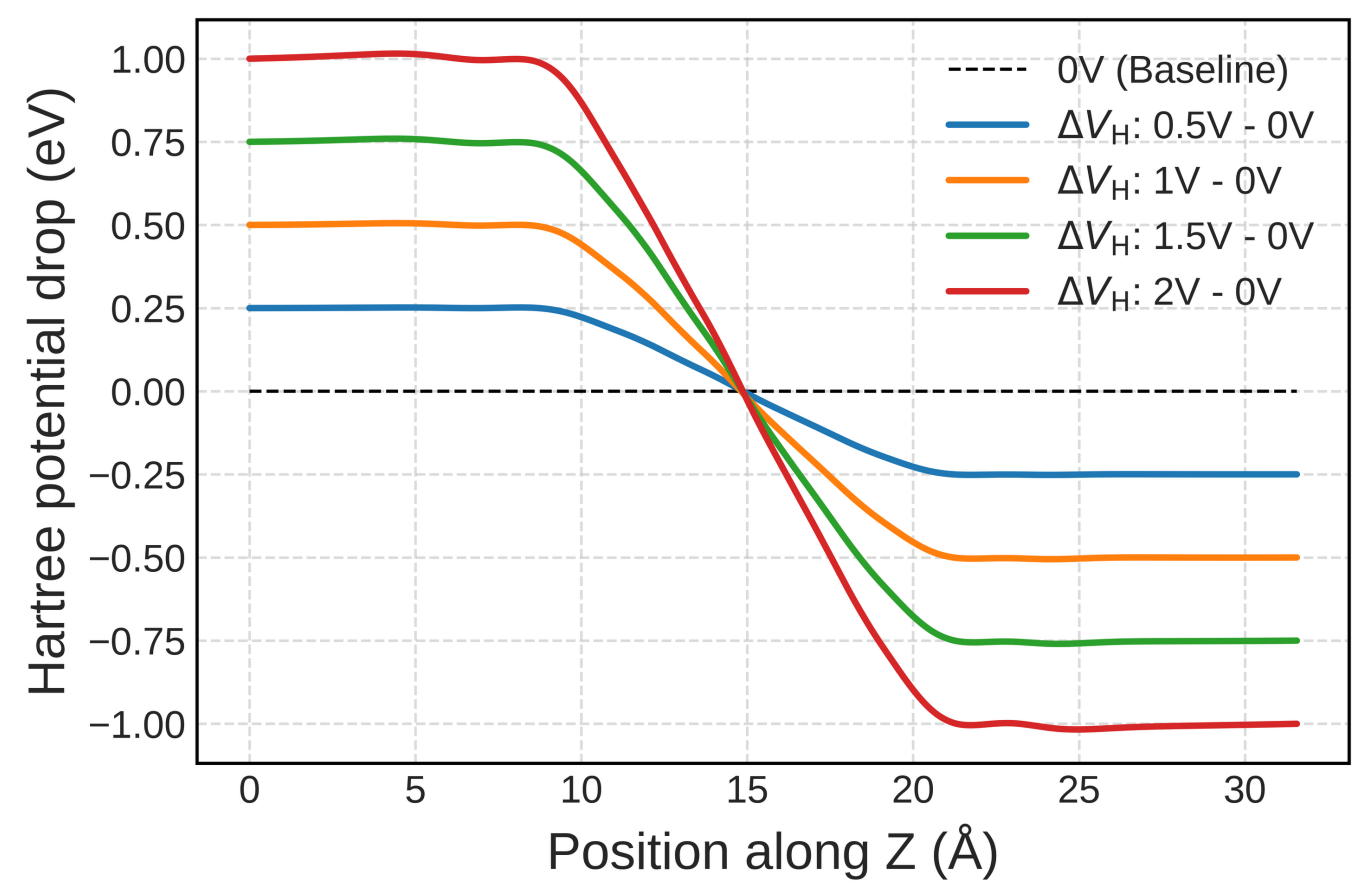

**Figure S9** Hartree potential drop profile ($\Delta V_H = V_{bias} - V_{0V}$) along $z$-axis.

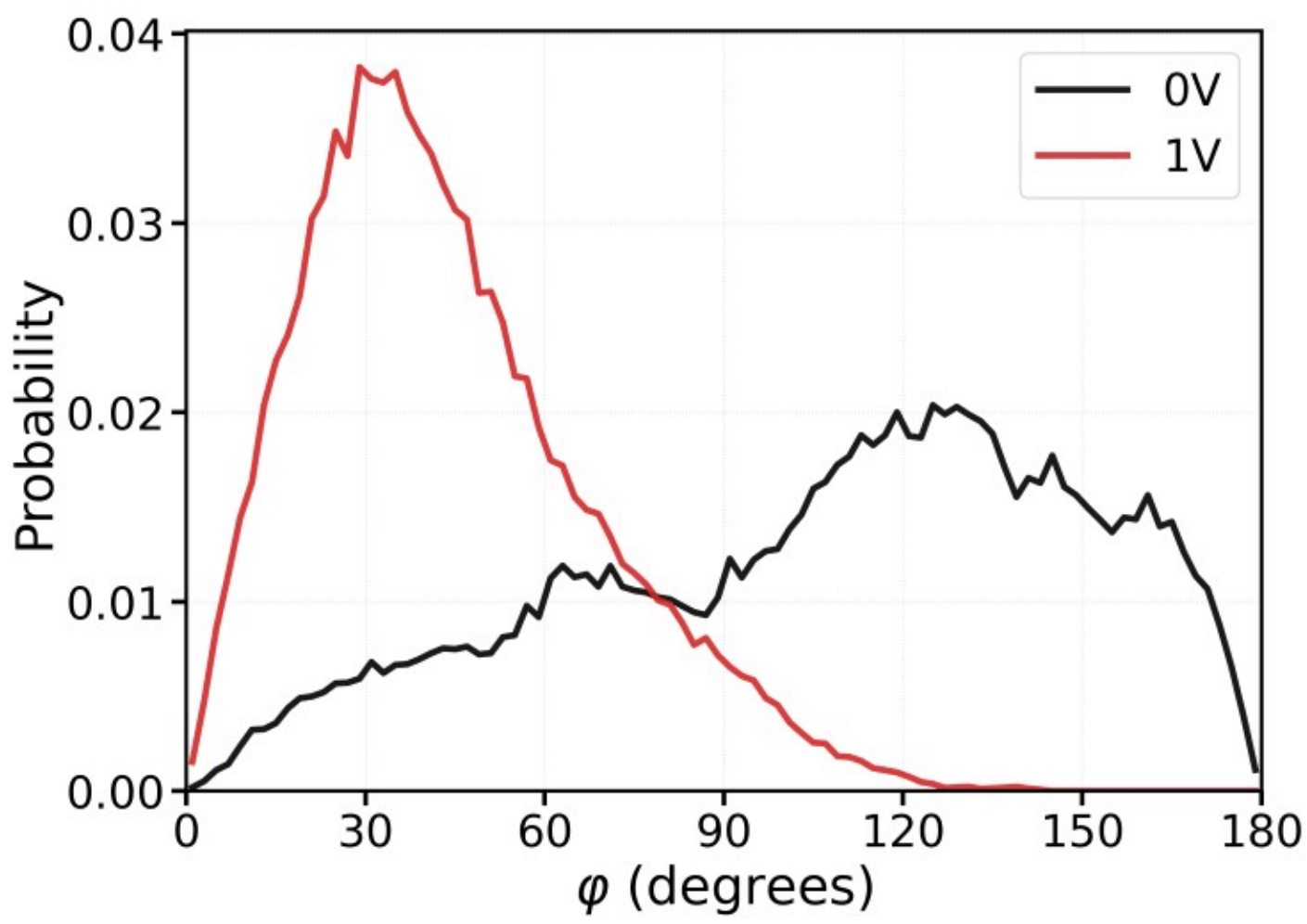

**Figure S10** Probability distribution of the angle $\varphi$ at 0 V and 1 V derived from AIMD.

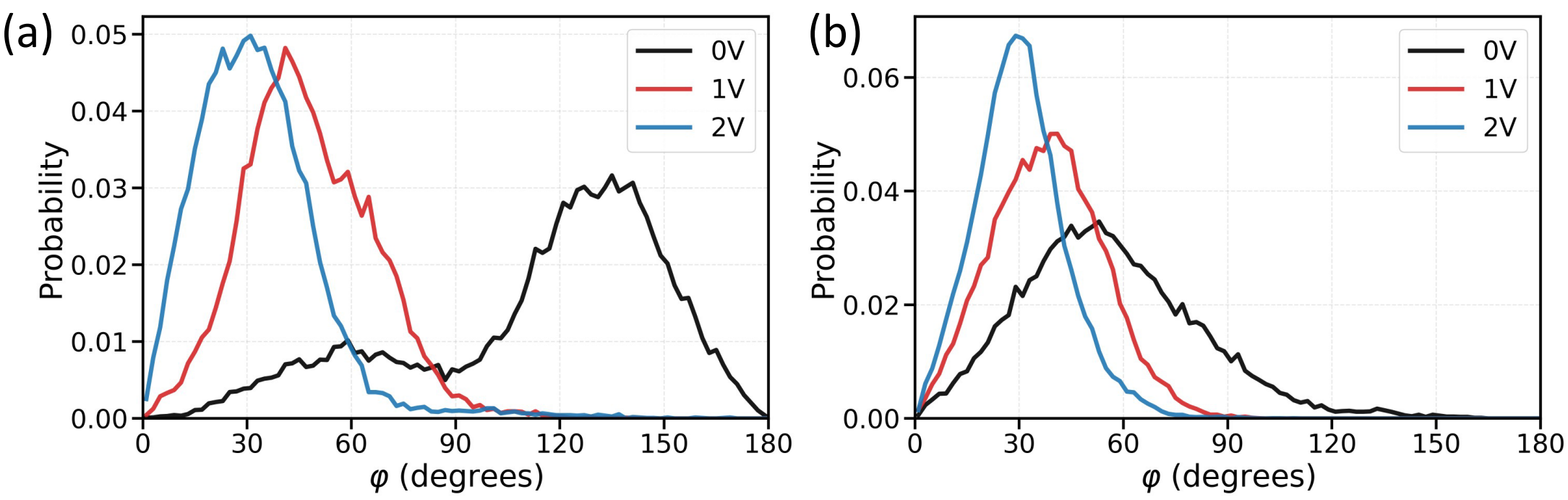

**Figure S11** Probability distribution of the angle $\varphi$ at 0 V, 1 V, and 2 V near **a,** left-lead and **b,** right-lead regions derived from ML-MD.

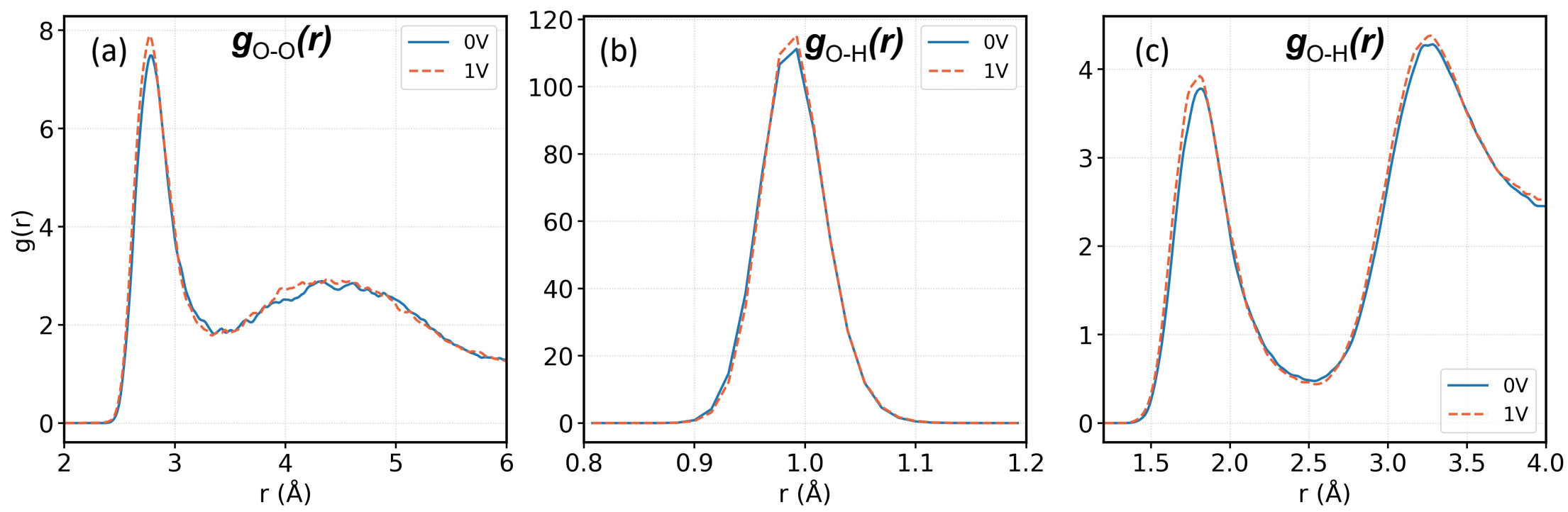

**Figure S12** Radial Distribution Functions (RDFs) calculated from AIMD trajectories: **a,** The O-O RDF $g_{O\text{-}O}(r)$. **b,** The O-H RDF $g_{O\text{-}H}(r)$ ranging from 0.8 – 1.2 Å. **c,** The O-H RDF $g_{O\text{-}H}(r)$ ranging from 1.2 – 4 Å.

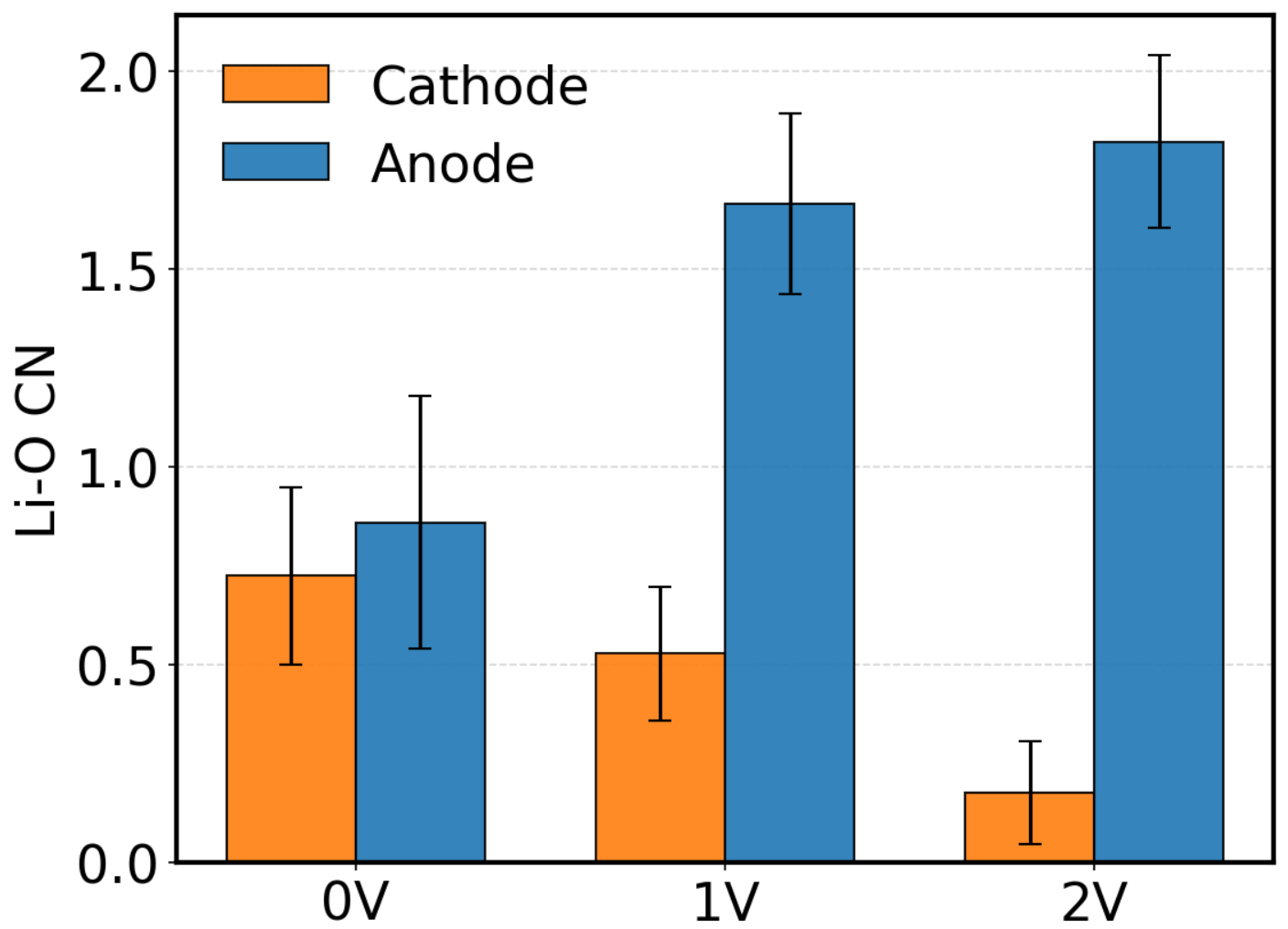

**Figure S13** Coordination number of Li–O at cathode and anode at different voltages. Only Li atoms at the lead surfaces (buffer region) are considered. The CN at cathode are 0.73, 0.53, 0.18 and at anode are 0.86, 1.67, 1.82, from 0 V to 2 V, respectively.

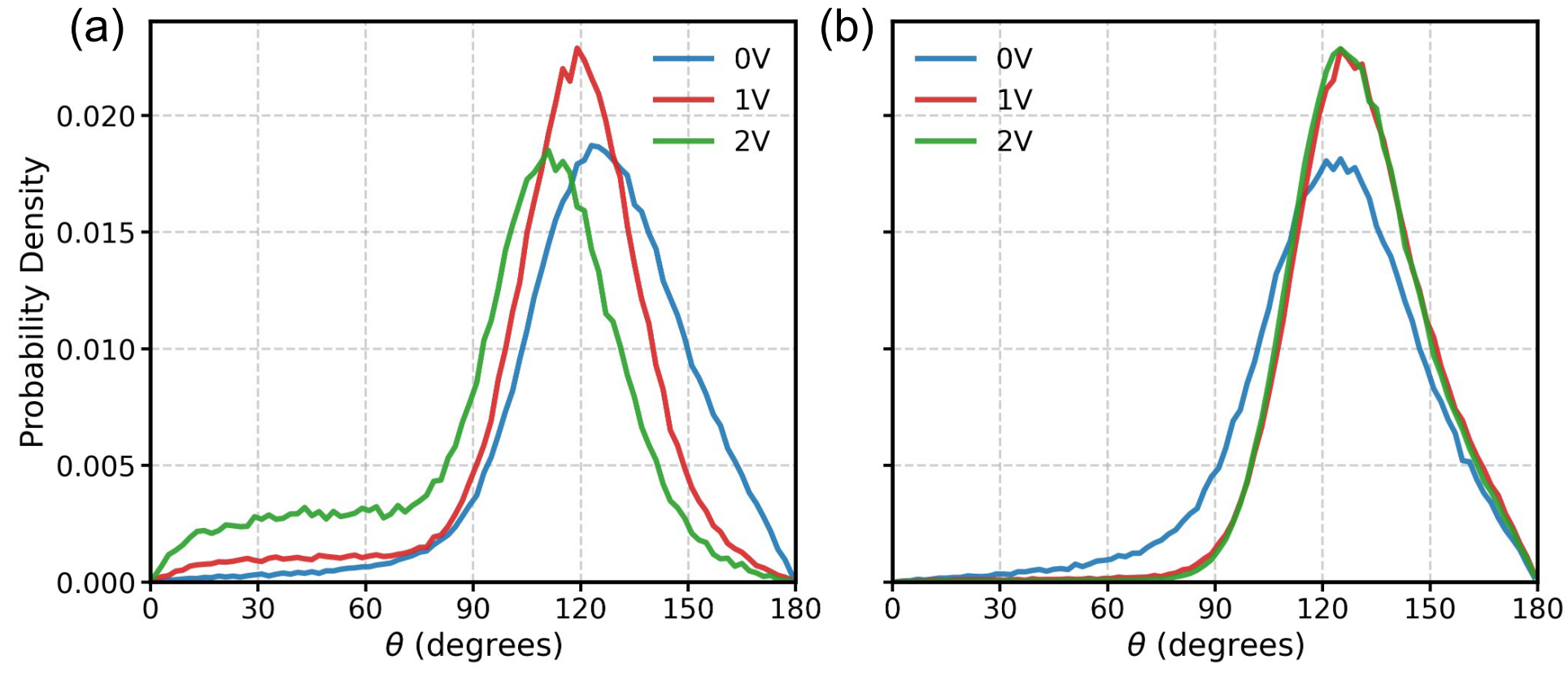

**Figure S14** Angular distribution of water dipoles relative to the Li-O vector ($\theta$) at the **a**, cathode and **b**, anode, respectively

## SI References